\documentclass[prd,aps,showpacs,preprintnumbers,amssymb]{revtex4}
\usepackage{graphicx}% Include figure files

\def\e3p{$\eta \rightarrow 3 \pi$}

\begin{document}

\title{%
\hfill{\normalsize\vbox{%
\hbox{\rm SU-4252-872}
 }}\\
{Note on a sigma model connection with instanton dynamics}}

\author{Amir H. Fariborz $^{\it \bf a}$~\footnote[3]{Email:
 fariboa@sunyit.edu}}

\author{Renata Jora $^{\it \bf b}$~\footnote[2]{Email:
 cjora@physics.syr.edu}}

\author{Joseph Schechter $^{\it \bf c}$~\footnote[4]{Email:
 schechte@physics.syr.edu}}

\affiliation{$^ {\bf \it a}$ Department of Mathematics/Science,
 State University of New York Institute of Technology, Utica,
 NY 13504-3050, USA.}

\affiliation{$^ {\bf \it b,c}$ Department of Physics,
 Syracuse University, Syracuse, NY 13244-1130, USA,}

\date{\today}

\begin{abstract}

    It is well known that the instanton approach to QCD generates
an effective term which looks like a three flavor determinant  
of quark bilinears. This has the right behavior to explain the
 unusual
mass and mixing of the $\eta(958)$ meson, as is often simply
 illustrated with the aid of a linear SU(3) sigma model. It is 
 less well known that the instanton analysis generates
 another term which has the same transformation property
but does not have a simple interpretation in terms of this usual
linear sigma model. Here we point out that this term has an
interpretation in a generalized linear sigma model
containing two chiral nonets. The second chiral nonet is
taken to correspond to mesons having two quarks and two antiquarks
in their makeup. The generalized model seems to be useful for 
learning about the spectrum of low lying scalar mesons which
have been emerging in the last few years. The physics of the new
term is shown to be related to the properties of an ``excited"
$\eta'$ state present in the generalized model and for which 
there are some experimental candidates.

\end{abstract}


\maketitle

\section{Introduction}

The instanton approach to QCD (see Refs.\cite{bpst}-\cite{th2} 
for some of the many interesting references) has played an 
important role in understanding the
origin of the $U(1)_A$ violation in that theory.
Specifically, 't Hooft showed \cite{th1} that a new
 quark term arises which conserves the $SU(N_f)_L$ x
$SU(N_f)_R$ symmetry, where $N_f$ denotes the number of relevant
low energy quark flavors, but violates $U(1)_A$. In the case
 $N_f$=2, the $U(1)_A$ violating term is a 2 x 2 determinant
of quark bilinears. If this is generalized to $N_f$=3, the 
resulting 3 x 3
determinant has the right transformation properties to explain
the unusually high mass as well as the mixing pattern of the
puzzling pseudoscalar meson, $\eta'$(958). The relevant 
calculation is often performed using an 
``effective low energy" linear SU(3) sigma model
 containing 
both a pseudoscalar nonet as well as an additional
scalar nonet. Such a model gives the usual ``precision"
current algebra results for the pion (and to some extent the
kaon) interactions and an acceptable description of the  
$\eta'$(958). It also contains information about the scalars
 although they are
 often ``integrated out". That
 procedure converts the 
model to a non-linear sigma model.

    It is amusing to note \cite{svz} that in the $N_f$=3 case,
the instanton calculation gives not only the determinant type
 $U(1)_A$ violation term but also another  $U(1)_A$ violation term
of non-determinant type. That term will be of interest in
 the present 
paper.
 We have been studying a generalized  linear sigma model
 \cite{BFMNS01}-\cite{3FJS07} containing two chiral nonets which
 are allowed to mix with each other.
Related models for thermodynamic properties of QCD
are discussed in Refs.\cite{ythb}.
 The underlying motivation
arises from the increasing liklihood \cite{inac} of the
 existence of 
light scalar mesons which show up
 for instance in the analysis of pion pion 
scattering data. Note that, at present, the scalars below 1 GeV
appear to fit into a nonet as:
\begin{eqnarray}
I=0: m[f_0(600)]&\approx& 500\,\,{\rm MeV}
\nonumber \\
I=1/2:\hskip .7cm m[\kappa]&\approx& 800 \,\,{\rm MeV}
\nonumber \\
I=0: m[f_0(980)]&\approx& 980 \,\,{\rm MeV}
\nonumber \\
I=1: m[a_0(980)]&\approx& 980 \,\,{\rm MeV}
\label{scalarnonet}
\end{eqnarray}
This level ordering is seen to be flipped compared to that
 of the standard vector 
meson nonet.
 It was pointed out a long time ago in Ref. \cite{j},
that the level order is automatically flipped
when mesons are made of two quarks and two antiquarks
instead of a single quark and antiquark. 
That argument was given for a diquark- anti diquark
structure but is easily seen to also hold for a meson-
meson, ``molecule" type structure which was advocated,
 at least for a
partial nonet,
 in Ref. \cite{wi}. Thus, on empirical grounds a
four quark structure for the light scalars seems 
plausible. Of course, one expects higher mass scalars related
  to p wave quark-antiquark composites to also exist.
 It is natural
to expect mixing between states with the same quantum
numbers and there is some phenomenological evidence for
this as noted in Refs \cite{BFS3} and \cite{mixing}. 
Thus, it seems reasonable to construct a
generalized linear sigma model containing a 
 chiral ``four quark" nonet as well as the usual
chiral ``two quark" nonet. The study of such a model
in fact yields a plausible explanation of
the main experimental facts. Of relevance to the 
instanton physics is that the two nonets are distinguished
from each other by having different $U(1)_A$ transformation 
properties. Furthermore, the treatment of the model in
\cite{1FJS07}-\cite{3FJS07} brings in an additional 
$U(1)_A$ violation term which seems to have the same structure
as the additional
term arising from the instanton analysis \cite{svz}.

    In section II, we repeat for the
 reader's convenience, the
notation \cite{BFMNS01} being used for 
schematic quark field combinations 
transforming
like chiral nonets with quark, antiquark and
various two quark, two antiquark structures.  

    In section III, we demonstrate that the
schematic molecule type chiral nonet can be written
as a linear combination of two diquark, antidiquark
type nonets (which have different SU(3) color 
representations for the diquarks). All of these ``four
quark" configurations have the same $U(1)_A$ transformation
property. Here it will be sufficient to assume
 that an unspecified
``four quark" configuration  is bound.     

In section IV, we give a brief outline of the 
linear sigma model containing both a ``two quark"
chiral nonet
and a ``four quark"
chiral nonet. It is convenient to introduce the
U(1)$_A$ violation in such a way that the classical 
Lagrangian mocks up the anomaly exactly. This leads
to ln's of the violation operators. It has the advantage
that the $\eta'$'s essentially decouple from the rest
of the particles. In order to compare with the instanton
analysis, we thus calculate the leading terms which are linear
in the violation operators.

In section V, we quote the known three flavor effective quark 
Lagrangian arising from the instanton analysis. We rewrite it
using Fierz transformations so that the desired
 ``four quark" fields
become manifest. They are presented
 as a linear combination of
of a ``molecule" type field and a field made
from a color ${\bar 3}$ diquark combined
with its corresponding anti diquark.

In section VI, we compare the relative
strengths of the two  U(1)$_A$ violation terms
as obtained from the instanton analysis to the ones
obtained from the generalized linear sigma model.
The linear sigma model relative
 term strengths are obtained from 
comparing the properties of the $\eta'$(958) with those
of the apparently best candidate to be its partner, the
$\eta$(1475). To convert the quark instanton Lagrangian to 
one involving only mesons we need a way to characterize our 
ignorance of the quark wave functions inside the meson states.
This is done via a parameter denoted $\omega$,
 which is estimated.

Section VII contains some concluding remarks.

\section{Notation}

   Even though one can not write down the exact QCD wave
 functions
of the low lying mesons it is easy to write down schematic
descriptions of how quark fields may combine to give particles
with specified transformation properties. For spinor notations
we employ the Pauli conventions. 
 We work in a representation where the $\gamma$
matrices and the charge conjugation matrix have the form:
\begin{eqnarray}
\gamma_i= \left[
\begin{array} {cc}
0&-i\sigma_i\\
i\sigma_i&0
\end{array}
\right], \hspace{1cm} \gamma_4=\left[
\begin{array}{cc}
0&1\\
1&0
\end{array}
\right],
\hspace{1cm} \gamma_5=\left[
\begin{array}{cc}
1&0\\
0&-1
\end{array}
\right],                    
\hspace{1cm}
 C= \left[
\begin{array} {cc}
-\sigma_2&0\\
0&\sigma_2
\end{array}
\right]. \label{candsigma}
\end{eqnarray}

A nonet $M(x)$ realizing the $q \bar q$ structure can be
written as:

\begin{equation}
M_a^b = {\left( q_{bA} \right)}^\dagger \gamma_4 \frac{1 +
\gamma_5}{2} q_ {aA}, \label{M}
\end{equation}
where $a$ and $A$ are respectively flavor and color indices. Our
convention
 for matrix notation is $M_a^b \rightarrow M_{ab}$.  Then $M$
 transforms under chiral $SU(3)_L \times SU(3)_R$, charge
 conjugation $C$ and 
parity $P$ as
\begin{eqnarray}
&&M \rightarrow U_L M U_R^\dagger
\nonumber\\
&&C: \quad M \rightarrow  M^T, \quad \quad P: \quad M({\bf x})
 \rightarrow  M^{\dagger}(-{\bf x}).
\label{Mchiral}
\end{eqnarray}
Here $U_L$ and $U_R$ are unitary, unimodular matrices associated
with the
 transformations on the left handed
 ($q_L = \frac{1}{2}\left( 1 + \gamma_5 \right) q$) and right
 handed ($q_R = \frac{1}{2}\left( 1 - \gamma_5 \right) q$)
 quark projections.
 For the $U(1)_A$ transformation one has:
\begin{equation}
M \rightarrow e^{2i\nu} M. \label{MU1A}
\end{equation}
Next consider nonets with `` four quark",
 $qq{\bar q}{\bar q}$ structures. 
One possibility is that the four quark states 
 are ``molecules''
 made out of two quark-antiquark states. This leads to
 the following schematic form:
\begin{equation}
M_a^{(2)b} = \epsilon_{acd} \epsilon^{bef}
 {\left( M^{\dagger} \right)}_e^c {\left( M^{\dagger}
 \right)}_f^d.
\label{M2}
\end{equation}

 Another possibility 
 is that the four quark states may be bound states of a
diquark and an anti-diquark. There are two choices if the
 diquark is required to belong to a ${\bar 3}$ representation of
flavor SU(3). In the first case it belongs to a ${\bar 3}$
 of color
and is a spin singlet with the structure,
\begin{eqnarray}
L^{gE} = \epsilon^{gab} \epsilon^{EAB}q_{aA}^T C^{-1} \frac{1 +
 \gamma_5}{2} q_{bB}, \nonumber \\
R^{gE} = \epsilon^{gab} \epsilon^{EAB}q_{aA}^T C^{-1} \frac{1 -
\gamma_5}{2}
 q_{bB}.
\end{eqnarray}
 Then the matrix $M$ has the form:
\begin{equation}
M_g^{(3)f} = {\left( L^{gA}\right)}^\dagger R^{fA}. \label{M3}
\end{equation}
  In the second case
 the diquark belongs to a $6$ representation of color and has spin 1.
  It has the schematic chiral realization:
\begin{eqnarray}
L_{\mu \nu,AB}^g = L_{\mu \nu,BA}^g = \epsilon^{gab}
 q^T_{aA} C^{-1}
 \sigma_{\mu \nu} \frac{1 + \gamma_5}{2} q_{bB}, \nonumber \\
R_{\mu \nu,AB}^g = R_{\mu \nu,BA}^g = \epsilon^{gab}
 q^T_{aA} C^{-1}
 \sigma_{\mu \nu} \frac{1 - \gamma_5}{2} q_{bB},
\end{eqnarray}
where $\sigma_{\mu \nu} = \frac{1}{2i}
 \left[ \gamma_\mu, \gamma_\nu
\right]
 $.  The corresponding $M$ matrix has the form
\begin{equation}
M_g^{(4) f} = {\left( L^{g}_{\mu \nu,AB}\right)}^\dagger
 R^{f}_{\mu
\nu,AB},
\label{M4}
\end{equation}
where the dagger operation includes a factor
 ${(-1)}^{\delta_{\mu 4}
+
 \delta_{\nu 4}}$.The nonets $M^{(2)}$, $M^{(3)}$ and 
 $M^{(4)}$ transform
  like $M$
 under all of $SU(3)_L \times SU(3)_R$, $C$, $P$. Under $U(1)_A$
 they transform as:
\begin{equation}
M^{(2)} \rightarrow e^{-4i\nu} M^{(2)}.
\end{equation}
     It is seen that the $U(1)_A$ 
transformation distinguishes the ``four quark" from the 
``two quark" states.

\section{Different four quark structures}

  Now we will show
  that $M^{(2)}$, $M^{(3)}$ and $M^{(4)}$
 are related by a Fierz
 transformation; thus only two of them are
 linearly independent. For this purpose it is convenient
to express the four component spinors in terms of the two
component chiral projections in the basis given above:
\begin{eqnarray}
q_{aA}=
 \left[
\begin{array} {c}
q_{LaA}\\
q_{RaA}
\end{array}
\right].
\label{LR}
\end{eqnarray}
The quark-antiquark field,
$M$ has the schematic structure $M_a^b =q^{\dagger}_{RbA}
q_{LaA}$ while 
  $(M^{\dagger})_b^a =q^{\dagger}_{LaA}
q_{RbA}$.                                             
Similarly,
the schematic molecule-type field $M^{(2)}$ takes the form:
\begin{eqnarray}
M_g^{(2)f}=
\epsilon_{gab}\epsilon^{fde}
\left[q^\dagger_{LaA}q_{RdA}\right]
\left[q^\dagger_{LbB}q_{ReB}\right].
\label{m2}
\end{eqnarray}
Using the definition $(\sigma_2)_{\alpha\beta}=
-i\epsilon_{\alpha\beta}$ and the anti-commutativity
of the fermi fields we readily obtain the decomposition
of $M^{(3)}$ as,
\begin{equation}
M^{(3)f}_g=2\epsilon_{gab}\epsilon^{fde}\left(
\left[q^\dagger_{LaA}q_{RdA}\right]
\left[q^\dagger_{LbB}q_{ReB}\right]-
\left[q^\dagger_{LaA}q_{RdB}\right]
\left[q^\dagger_{LbB}q_{ReA}\right]\right).
\label{m3}\end{equation}
To simplify $M^{(4)}$ we make use of the well
 known identity
$\sigma_2\sigma_k^*\sigma_2=-\sigma_k$
and also the Fierz type relation,
\begin{equation}
(\sigma_k\sigma_2)_{\beta\alpha}
(\sigma_2\sigma_k)_{\eta\rho}
=\delta_{\beta\rho}\delta_{\alpha\eta}+
\delta_{\beta\eta}\delta_{\alpha\rho}.
\label{sigmaid}
\end{equation}
Then we find,
\begin{equation}
M^{(4)f}_g=-4\epsilon_{gab}\epsilon^{fde}\left(
\left[q^\dagger_{LaA}q_{RdA}\right]
\left[q^\dagger_{LbB}q_{ReB}\right]+
\left[q^\dagger_{LaA}q_{RdB}\right]
\left[q^\dagger_{LbB}q_{ReA}\right]\right).
\label{m4}
\end{equation}                          
Now it is easy to see that the molecule-type
field $M^{(2)}$
may be expressed as a linear
combination of $M^{(3)}$ and $M^{(4)}$:
\begin{equation}
M_a^{(2)b}=\frac{2M_a^{(3)b}-M_a^{(4)b}}{8}.
\label{lineardep}
\end{equation}
Thus, at a naive quark model level, there
is no absolute distinction between the molecule
type field and a linear combination of two
different diquark-antidiquark configurations.
It may be amusing to note that, in the MIT bag
model approach \cite{mit} to four quark scalars,
 the relevant eigenstates of the 
hyperfine splitting Hamiltonian also emerge as
a linear combination
of two
 diquark-antidiquark configurations.
 Of course there may be
differences which would emerge
if the full QCD dynamics could be solved. 
Some dynamical arguments are discussed in Ref.
 \cite{kknhh05}.

    There are no external quantum numbers to differentiate
 $M^{(2)}$,  $M^{(3)}$,
and  $M^{(4)}$ from each other. Thus we just assume that
the dynamics selects  a particular but unknown
 linear combination of (any two of) them to be a bound 
``four quark" field,
$M^{\prime}$. Note, however, that $M$ and $M^{\prime}$
 are distinguished from
each other by their different $U(1)_A$ transformation properties.

\section{Effective potential}
In our model Lagrangian we use scalar fields with the 
transformation properties of the schematic fields $M$ and $M'$
just discussed.
These fields may be decomposed into hermitian scalar (S) and
pseudoscalar ($\phi$) nonets as,
\begin{eqnarray}
M &=& S +i\phi, \nonumber  \\
M^\prime &=& S^\prime +i\phi^\prime. \label{sandphi}
\end{eqnarray}
The Lagrangian density for our model is taken to have
the simple form,
\begin{equation}
{\cal L} = - \frac{1}{2} {\rm Tr}
 \left( \partial_\mu M \partial_\mu M^\dagger
 \right) - \frac{1}{2} {\rm Tr}
 \left( \partial_\mu M^\prime \partial_\mu M^{\prime \dagger}
 \right)
 - V_0 \left( M, M^\prime \right) - V_{SB},
\label{mixingLsMLag}
\end{equation}
with non-derivative interaction terms.
Here $V_0(M,M^\prime) $ stands for a general function made
 from $SU(3)_L \times SU(3)_R$ but not necessarily $U(1)_A$
 invariants formed out of $M$ and $M^\prime$.
 Furthermore $V_{SB}$
is a flavor symmetry breaking term designed to model
the quark mass terms in QCD.

    Generally one has the situation
 where non-zero vacuum values
of the diagonal components of $S$ and $S'$
 may exist. These will be
denoted by,
\begin{equation}
\left< S_a^b \right> = \alpha_a \delta_a^b,
 \quad \quad \left< S_a^{\prime b} \right> =
\beta_a \delta_a^b. \label{vevs}
\end{equation}
In the iso-spin invariant limit, $\alpha_1=\alpha_2$ and
$\beta_1=\beta_2$ while in the SU(3) invariant limit,
$\alpha_1=\alpha_2=\alpha_3\equiv \alpha$
 and $\beta_1=\beta_2=\beta_3\equiv\beta$.

     The model is an upgrading of the single-M SU(3)
linear sigma model to one containing two chiral nonets.
However, it is much more complicated. For example,
 the renormalizable version of the present model has
(see Appendix A of \cite{FJS05} and Appendix A of
\cite{2FJS07} ) twenty one invariant terms in $V_0$ while the 
renormalizable version of the single-M model has
 only four terms. To make progress we suggested 
  first including only those terms with no more
than a total of eight (quark plus antiquark) lines
in the underlying schematic interaction. This
led to the predictions(\cite{2FJS07}, \cite{3FJS07})
:i) a very light singlet
scalar which might be identified with the $f_0$(600), 
ii) large four quark content of the lighter scalars, 
iii) improved s-wave pion pion isosinglet scattering length.
The model included two $U(1)_A$ violating, but
chiral SU(3) conserving, terms. These were chosen
to mock up the $U(1)_A$ anomaly of QCD. That is a
reasonable requirement in the present context
since the $U(1)_A$ symmetry distinguishes the
two quark from the four quark mesons.

   In the single-M model, it was noted
(see Appendix of \cite{su71}) even before QCD,
that a determinant type U(1)$_A$ violating piece was needed
 to explain the
 $\eta$ mesons. After
QCD, 't Hooft \cite{th1} showed that a quark level term
of the required sort would arise from instanton
contributions. Actually, he did not completely present the 
relevant three flavor version of his model.
 Other authors \cite{svz}
later gave this result and one can see that there is an
additional  U(1)$_A$ violation term present. Here we will
 show that the additional term has 
the same structure as the one we added on the basis of
the quark counting just mentioned.
    
    In the effective Lagrangian framework the
  axial anomaly was first ``exactly" modeled
\cite{U1A} 
by including a term 
proportional to $G({\rm ln} {\rm det} M - 
{\rm ln} {\rm det} M^\dagger)$,
where G represents the pseudoscalar Yang Mills
invariant ${\rm Tr} (F_{\mu\nu}{\tilde 
F}_{\mu\nu})$,
constructed from the field strength tensor.
 It is necessary to include a wrong
 sign mass term for G which is then 
integrated out. Then one obtains a form like
\begin{equation}
{\cal L}_\eta =-c_3
\left[{\rm ln} 
\left(\frac{{\rm det} 
M}{{\rm det} M^{\dagger}}
\right)\right]^2,
\label{etaprmass}
\end{equation}
where $c_3$ is a numerical parameter. In the present
 model with two
chiral nonets this form is not unique and the most plausible
modification \cite{1FJS07} is to replace 
${\rm ln} (\frac{{\rm det} M}{{\rm det}
M^{\dagger}})$ by
\begin{equation}
\gamma_1
\left[{\rm 
ln}
\left(
\frac{{\rm 
det} (M)}{{\rm det} (M^\dagger)}
\right)\right]+(1-\gamma_1)
\left[{\rm ln} 
\left(
\frac{{\rm Tr}(MM'^\dagger)}{
{\rm Tr} (M'M^\dagger)} 
\right)
\right],
\label{gamma1}
\end{equation}
where $\gamma_1$ is a dimensionless parameter. For the
purpose of comparison with instanton results in the next
section we will approximate this somewhat complicated form
by its leading term. With the assumption that 
$\frac{{\rm det}(M)}
{\langle {\rm det}(M)\rangle}= 1 + small$, we 
write: 
\begin{equation}
{\rm ln}\left[ {\rm det} 
(M)\right] \approx
\left\langle {\rm 
det} (M) \right\rangle +
\left[\frac{{\rm det}(M)}
{\langle {\rm det} (M)\rangle}-1\right].
\label{app1}
\end{equation}                            
Then,
\begin{eqnarray}
&&
\left( {\rm ln}[ {\rm det} (M) ] - {\rm ln} 
[{\rm det} (M^\dagger)]
\right)^2 \approx
\left[\frac{{\rm 
det} (M)-{\rm det} (M^\dagger)}{\langle 
{\rm det} (M) \rangle}
\right]^2
\nonumber \\
&&=\frac{1}{\alpha^6}[({\rm 
det} (M)+ {\rm det} (M^\dagger))^2-
4 {\rm det} (MM^\dagger)]\approx 
\frac{4}{\alpha^3}
({\rm det} (M)+ {\rm det}(M^\dagger)).
\label{app2}
\end{eqnarray}
In this procedure a purely numerical constant has been dropped
and the $U(1)_A$ invariant piece, 
${\rm det}(MM^\dagger)$ was
considered small compared to other $U(1)_A$ invariant pieces.
Similarly,
\begin{equation}
\left[
{\rm ln}\left(
\frac{{\rm Tr} (MM'^\dagger)} 
{{\rm Tr} (M'M^\dagger}\right)	
\right]^2 \approx 
\frac{4}{3\alpha\beta}
[{\rm Tr} (MM'^\dagger) + {\rm 
Tr} (M'M^\dagger)].
\label{app3}
\end{equation}
Cross terms from squaring Eq.(\ref{gamma1})
are neglected in the same
approximation.
Summarizing these steps we write,
\begin{eqnarray}
{\cal L}_\eta &=&-c_3\left(
\gamma_1 
{\rm 
ln} 
\frac{{\rm 
det}(M)}{{\rm det}(M^\dagger)}+(1-\gamma_1)
{\rm ln} \frac{{\rm Tr} (MM'^\dagger)}{
{\rm Tr} (M'M^\dagger)}\right)^2
\nonumber  \\
&\approx&-4c_3\left(\frac{\gamma_1^2}{\alpha^3}
[{\rm det}(M)+ {\rm det}(M^\dagger)]+
\frac{(1-\gamma_1)^2}{3\alpha\beta}
[{\rm Tr} 
(MM'^\dagger)+ {\rm Tr}(M'M^\dagger)]\right).
\label{lastapprox}
\end{eqnarray}
In contrast to these approximations,
 keeping the ln's in the
 calculations involving the mesonic Lagrangian,                                              
 leads to a desirable simplifying
 decoupling of the $\eta'$ sector
 of the model from the parts which conserve $U(1)_A$.
 This was previously discussed in 
\cite{1FJS07}-\cite{3FJS07}.

\section{U(1)$_A$ violation from instantons}

    The 't Hooft effective Lagrangian for the three flavor
case \cite{svz} can be presented \cite{presentation} as:
\begin{eqnarray}
{\cal L}(q)&&=const \frac{1}{6 N_c(N_c^2-1)}
\epsilon_{gab}
\epsilon^{fde}
\nonumber \\
&&
\left(
\frac{2N_c+1}{2N_c+4}
\left[{\bar q}_{gA}\frac{1+\gamma_5}{2}
q_{fA}\right]
\left[{\bar q}_{aB}\frac{1+\gamma_5}{2}
q_{dB}\right] 
\left[{\bar 
q}_{bC}\frac{1+\gamma_5}{2}
q_{eC}\right]
\right.
\nonumber\\
&&+\frac{3}{8(N_c+2)}
\left[
{\bar q}_{gA}\frac{1+\gamma_5}{2}
q_{fA}\right]
\left[{\bar 
q}_{aB}\frac{1+\gamma_5}{2}\sigma_{\mu\nu}
q_{dB}\right] 
\left[{\bar 
q}_{bC}\frac{1+\gamma_5}{2}\sigma_{\mu\nu}
q_{eC}\right]
\nonumber\\                                    
 &&
\left.+
\left[\frac{1+\gamma_5}{2}\rightarrow\frac{1-\gamma_5}{2}\right]
\right).
\label{effL}
\end{eqnarray}                                                           
Here the kinematics were modified from Euclidean
space, appropriate for the path integral derivation, to
ordinary Minkowski space. The overall constant contains
a function of the QCD running coupling constant which
essentially cuts it off at higher energies.

 The quantities like $[{\bar q}_{gA}\frac{1+\gamma_5}{2}
q_{fA}]$ which appear in this equation clearly can be identified
with the usual quark antiquark 
meson field $M_f^g$ defined in Eq.(\ref{M}). The quantities
involving $\sigma_{\mu\nu}$ on the third line are less familiar.
Using the identity,
\begin{equation}
(\sigma_k)_{\beta\alpha}
(\sigma_k)_{\eta\rho}
=2\delta_{\beta\rho}\delta_{\alpha\eta}-
\delta_{\beta\alpha}\delta_{\eta\rho}.   
\label{commid}
\end{equation}
we find
\begin{eqnarray}
&&\epsilon_{gab}\epsilon^{fde}
\left[{\bar 
q}_{aB}\frac{1+\gamma_5}{2}\sigma_{\mu\nu}
q_{dB}\right] 
\left[{\bar 
q}_{bC}\frac{1+\gamma_5}{2}\sigma_{\mu\nu}
q_{eC}\right]                       
\nonumber \\
&&=4\epsilon_{gab}\epsilon^{fde}\left(
2
\left[q^\dagger_{RaB}q_{LdC}\right]
\left[q^\dagger_{RbC}q_{LeB}\right]
-\left[q^\dagger_{RaA}q_{LdA}\right]
\left[q^\dagger_{RbD}q_{LeD}\right]\right)
\nonumber \\
&&=4\left[(M^{(2)\dagger})^f_g-(M^{(3)\dagger})^f_g
\right],
\label{sigmunu}                                       
\end{eqnarray}
where Eqs.(\ref{m2}), (\ref{m3}) and (\ref{lineardep})
  were used in the last step.
Putting these identifications
 back into Eq.(\ref{effL}) finally yields:
\begin{equation}
{\cal L}(q)= \frac{const}{2N_c(N_c^2-1)(N_c+2)} 
\left[(2N_c+1) 
{\rm 
det}(M(q))+\frac{1}{2}{\rm 
Tr} \left[M(q)(M^{(2)\dagger}(q)
-M^{(3)\dagger}(q))\right]
\right] + h.c.
\label{newform}
\end{equation}
Here the determinant and trace refer to the
 three-flavor space.

\section{Comparison of sigma model and instanton approaches}

    It is immediately clear that the $U(1)_A$ violating instanton 
generated Lagrangian of Eq.(\ref{newform})
 has the same structure as
${\cal L}_{\eta}$, the
linearized $U(1)_A$ violating Lagrangian
 in Eq.(\ref{lastapprox}). Of 
course,
the sigma model expression is constructed out of physical meson
fields while the instanton expression is constructed out of
 schematic combinations of quark fields with the same
 transformation 
properties. Presumably the schematic quark combinations
 will be dominated by, 
or at least have substantial overlap with, the
corresponding meson fields. This similarity seems to be 
the strongest 
point of our discussion. It is especially interesting to us
in the context of building linear sigma models to learn
 about possible
mixing of quark-antiquark and two quark plus two antiquark mesons.
As noted in section IV, even the {\it renormalizable}
linear sigma model  potential would have  
 too many terms for practical analysis.
We therefore suggested a simplifying scheme
in which terms with the smallest number of underlying
(quark plus antiquark) fields be retained.
On this basis, the two dominant  $U(1)_A$ 
violating terms
are expected to be the ${\rm det}(M)$ and 
${\rm Tr} (MM'^\dagger)$ ones, each 
representing six underlying fermions.
 This is apparently confirmed by the leading
instanton calculation. The four quark 
structure appearing in Eq.(\ref{newform}) is
seen to contain $M^{(2)}-M^{(3)}$, a linear 
combination with 
equal strengths of "molecule" type and diquark plus anti diquark
components. This does not guarantee, naturally, that such
 a combination
is the one which is dynamically bound.

    To get a rough indication of what is happening, we introduce
a prescription for obtaining the leading $U(1)_A$ violating  
terms in the meson Lagrangian given in Eq.(\ref{lastapprox});
we simply make
the replacements,
\begin{eqnarray}
&&M(q)\rightarrow-\Lambda^2M,
\nonumber \\
&&M^{(2)}(q)-M^{(3)}(q)\rightarrow\pm \omega \Lambda^5 M',
\label{prescrip}
\end{eqnarray}
in the instanton Lagrangian of Eq.(\ref{newform}).
Here, $M$ and $M'$ are the meson fields while $M(q)$ etc.
represent corresponding schematic quark structures with the same
chiral transformation properties. 
The positive quantity $\Lambda$ is a QCD type scale with 
dimension of mass.
The dimensionless, positive quantity 
 $\omega$ is a phenomenological parameter introduced to
  account for our ignorance of which linear combination
 of the possible
 four quark states is actually bound, 
 the possibility
 of other hadronized field combinations appearing
 in Eq.(\ref{sigmunu}) as
well as other QCD effects. Finally the sign choice
 measures the sign of
the vacuum value of the left hand side operator.

   We may estimate $\Lambda$ above by taking the
ground state expectation value of the (11) matrix element:
\begin{equation}
\langle M_{11} \rangle = \alpha = \frac{-1}{2\Lambda^2}
\langle {\bar q}_{1A}q_{1A}\rangle.
\label{getlam}
\end{equation} 
Using $\alpha=0.0606$ GeV,
 as obtained in the SU(3) limit for either 
the massless or massive pion cases in \cite{2FJS07}
  or \cite{3FJS07} 
together with $\langle {\bar q}_{1A}q_{1A}\rangle \approx$ - 0.016
GeV$^3$ yields for the scale factor, $\Lambda \approx$ 0.36 GeV. 
As a check on this procedure, if the ``quark mass" factor, $A$
in the symmetry breaking term ${\cal 
L}_{SB}=2{\rm Tr} (AS)$ is
also scaled
as $A=\Lambda^2 m_q$, where $m_q$ is the diagonal matrix of quark 
masses, then the current algebra mass formula for the
 pseudoscalars
is converted to the corresponding linear sigma model
mass formula.

\begin{figure}[htbp]
\centering
%\rotatebox{270}
{\includegraphics[width=7cm,clip=true]{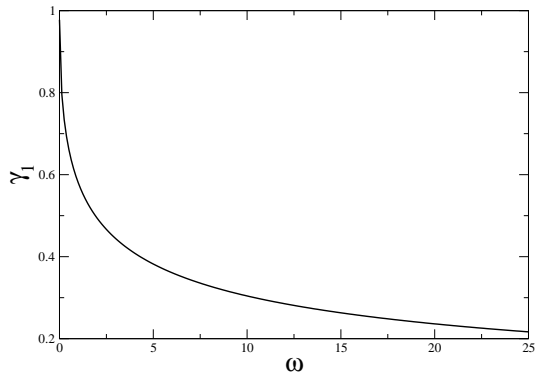}}
\caption[]{
$\gamma_1$ vs. $\omega$ for the positive sign choice
in Eq.(\ref{gamma1Q}) 
}
\label{gvo1}
\end{figure}

\begin{figure}[htbp]
\centering
%\rotatebox{270}
{\includegraphics[width=7cm,clip=true]{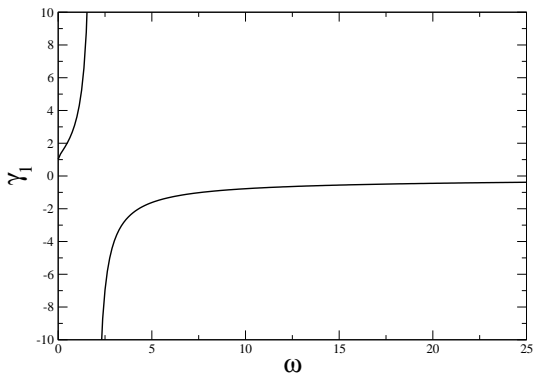}}
\caption[]{
$\gamma_1$ vs. $\omega$ for the negative sign choice
in Eq.(\ref{gamma1Q}) }
\label{gvo2}
\end{figure}

  Notice that the overall factor in Eq.(\ref{newform}) is heavily 
suppressed for large $N_c$ as expected for instanton effects. Thus
we shall set $N_c$ to be three, for
 our world. Then the substitutions
of Eq.(\ref{prescrip}) result in a meson U(1)$_A$ violating
Lagrangian of the form:
\begin{equation}
{\cal L}_\eta= const'\left[-7 
{\rm det}(M)\mp\frac{\omega\Lambda}{2}
{\rm Tr} (MM'^{\dagger})\right] +h.c.
\label{compare}
\end{equation}
Comparing Eq.(\ref{compare}) with Eq.(\ref{lastapprox})
gives a relation between $\gamma_1$,
a measure of the relative
strengths of the det and Tr terms which are being used to 
model the $U(1)_A$ anomaly, and the scaling factor $\omega$
introduced in Eq.(\ref{prescrip}):
\begin{equation}
\frac{\gamma_1^2}{(1-\gamma_1)^2}=
\frac{\pm 14\alpha^2}{3\beta\Lambda\omega}.
\label{gamma1omega}
\end{equation}
Since the left hand side of this equation must be positive
and $\beta (\approx$
0.0249 GeV) is positive, consistency
 requires us to keep only the + sign
in Eq.(\ref{prescrip}).
 This corresponds
to a positive vacuum value for $M^{(2)}(q)-M^{(3)}(q)$,
as opposed to the negative vacuum value for $M(q)$
shown in Eq.(\ref{getlam})
Defining
$Q(\omega)=14\alpha^2/(3\beta\Lambda\omega)$, this
quadratic equation has the solutions
\begin{equation}
\gamma_1(\omega)=\frac{-Q\pm{\sqrt Q}}{1-Q}.
\label{gamma1Q}
\end{equation}
 Note that the $\pm$ sign here is related to 
solving the quadratic equation rather
than to the possible choice displayed in 
Eq.(\ref{prescrip}). Fig.\ref{gvo1} shows that,
for the + sign choice, 
$\gamma_1(\omega)$ is positive and slowly
decreasing as $\omega$ increases.
On the other hand for the - sign choice,
$\gamma_1(\omega)$ can be either positive
or negative, as seen in Fig.\ref{gvo2}. For either
sign choice $\gamma_1(\omega)$ goes to one as
$\omega$ goes to zero.

%\begin{figure}[h]
%\begin{center}
%\vskip 1cm
%\epsfxsize = 7.5cm
%\ \epsfbox{gavsom1.eps}
%\end{center}
%\caption[]{
%$\gamma_1$ vs. $\omega$ for the positive sign choice
%in Eq.(\ref{gamma1Q}) }
%\label{gvo1}
%\end{figure}

%\begin{figure}[h]
%\begin{center}
%\vskip 1cm
%\epsfxsize = 7.5cm
%\ \epsfbox{gavsom2.eps}
%\end{center}
%\caption[]{
%$\gamma_1$ vs. $\omega$ for the negative sign choice
%in Eq.(\ref{gamma1Q}) }
%\label{gvo2}
%\end{figure}

    In our mesonic level sigma model the value of 
 $\gamma_1$ affects the masses and mixings of the  
four pseudoscalar isosinglets which appear. The
lowest lying ones are the $\eta$(547) and the
 $\eta$(958) while there are four  experimental candidates
(with masses in MeV at 1295, 1405, 1475 and 1760) for the
two higher lying states. This is, in general,
a complicated mixing problem with some experimental
ambiguities. In Refs \cite{2FJS07}(see section IV
and Appendix B) and \cite{3FJS07}(see Appendix A)
 we examined 
flavor SU(3) symmetric situations and considered the
favored scenario to be the one 
in which the $\eta$(958) mixed with 
the $\eta$(1475). Furthermore there
were two possible solutions with different
``four quark contents".
 The preferred solution with mainly 
``two quark content" for the $\eta'$
[denoted I2 in Appendix B of  \cite{2FJS07}] gave 
$\gamma_1\approx$ 0.25 while the somewhat less favored
solution [labeled I1] with less ``two quark content" for the 
 $\eta'$ gave $\gamma_1\approx$ 0.54. These
 are seen to
result respectively in values of about 18 and 1.3 for
$\omega$. Clearly, $\omega$ is sensitive to
the value of  $\gamma_1$, although it might be fairer to 
compare the values of $\omega^{1/5}$ which differ less
for the two alternatives.

     It may be interesting to present these results in terms of 
quantities associated with the quark level instanton Lagrangian
in Eq.(\ref{effL}). Using Eq.(\ref{prescrip}) and comparing with
 Eq.(\ref{lastapprox}) gives for the overall constant:
\begin{equation}
const=\frac{960c_3\gamma_1^2}{7\alpha^3\Lambda^6}.
\label{const}
\end{equation}
Similarly, the vacuum value of $M^{(2)}(q)-M^{(3)}(q)$ can
be estimated as:
\begin{equation}
\langle M'(q)\rangle = 
\langle [M^{(2)}(q)-M^{(3)}(q)]_{11}\rangle
=\omega\beta\Lambda^5.
\label{vev2}
\end{equation}
The overall constant and the ``four quark" vacuum value are listed in 
Table \ref{table} for each of the two considered scenarios. For 
comparison, the square of the ``two quark" vacuum value given in 
Eq.(\ref{getlam}) is 3.7 $\times 10^{-3}$ GeV$^6$,
 similar in order of 
magnitude to the four quark ones.

\begin{table}[htbp]
\begin{center}
\begin{tabular}{c|c|c}
\hline \hline
                   &       I1             & I2\\
$c_3 ({\rm GeV}^4)$&$-2.42 \times 10^{-4}$&$-2.42
 \times 10^{-4}$\\
$\gamma_1$         & 5.4 $\times 10^{-1}$ &  $2.5
 \times 10^{-1}$\\
$const ({\rm GeV}^{-5})$            &$-1.99\times 
10^{4}$&$-0.43\times 10^{4}$\\
$\langle M'(q) \rangle  ({\rm GeV}^{6})$            &$1.96 
\times 10^{-4}$&$2.7 \times 10^{-3}$\\
\hline
\end{tabular}
\end{center}
\caption[]{Estimated values of the overall constant,
$const$ and the vacuum value, $\langle M'(q)\rangle$
for the quark level Lagrangian.}    
 \label{table}
\end{table}

 Of course, the
results just discussed will be modified to some extent by
the inclusion of SU(3) flavor symmetry breaking effects. This 
work, which is under study, involves at minimum the consideration
of a 4 x 4 mixing matrix for the isoscalar pseudoscalar
 mesons in the 
present framework. 
  
 \section{Summary and discussion}
 
    We have shown that an extra term in the effective instanton 
generated Lagrangian has a natural interpretation as a mixing
term between quark-antiquark spin zero mesons and spin zero mesons
made from two quarks and two anti-quarks (in some unspecified
combination). Since the fields of the
 two kinds of mesons have different
U(1)$_A$ quantum numbers (before quark masses and
 spontaneous symmetry
symmetry breaking are taken into account) this term
also  violates the U(1)$_A$ symmetry.

    An interesting treatment of the relation between
instanton physics and the pattern of light scalar
meson decay widths has been
recently given in Ref. \cite{tsm}.

    On the question of what is the correct bound state
of ``four quark" mesons, we showed that at the zero quark mass
kinematical level the ``molecule" type could be rewritten
as a linear combination of two different diquark-antidiquark
 types.

    We worked at the level of a generalized linear SU(3) sigma
model which contains two scalar nonets and two
 pseudoscalar nonets.
The mixings between the two scalar nonets play
an important role in explaining the properties which seem
to be emerging from analysis of experimental data. The ``extra"
term of interest, on the other hand,
 primarily affects the mixing of the 
pseudoscalar SU(3) singlets. Indeed, we used a variation of
the model in which the axial anomaly was ``exactly" modeled,
 which has the effect of decoupling the pseudoscalar SU(3)
singlets. Using the masses of the $\eta'$(958)
and the $\eta$(1475) in the sigma model we made
 numerical estimates of the overall constant for the instanton
Lagrangian and the vacuum value of the ``four quark" operator     
which appears in it.

    The generalized linear sigma model in question is
actually a very complicated one, describing many different
particles and potentially having many different relevant terms.
Thus while, due to chiral symmetry,
 it gives a good description of
near threshold pion pion scattering for example,
 it is probably best regarded as a toy model for learning
 when it comes to describing a nonet's
worth of heavy pseudoscalars for example.

   From the point of view of truncating the terms
of this linear sigma model to a more manageable
number we had made \cite{1FJS07}-\cite{3FJS07}
 a provisional ansatz that terms
representing more than eight underlying fermion lines
be discarded. This gave two  U(1)$_A$
violating terms and 
corresponded nicely to the instanton
effective Lagrangian, which has two 
terms with six such lines. One might
ask about going beyond this approximation for an
effective model. If one allows 10 underlying fermion lines,
 the U(1)$_A$ violating terms with coefficients
 $e_3^b$, $e_4^d$ and $e_4^i$ in Eq.(A1) of \cite{FJS05} 
are possible. If one allows 12 underlying fermion lines
the terms with coefficients $d_3$ and $e_4^c$ kick in.
Finally, if one allows 14 underlying fermion lines the
terms with coefficients $e_4^e$ and $e_4^j$
 become possible. This could conceivably also be
an interesting expansion in the instanton approach.

\section*{Acknowledgments} \vskip -.5cm We are
happy to thank A. Abdel-Rehim, D. Black, M.
Harada, S. Moussa, S. Nasri and F. Sannino for
many helpful related discussions. 
The work of
A.H.F. has been partially supported by the NSF
Award 0652853.
The work
of R.J. and J.S. is supported in part by the U.
S. DOE under Contract no. DE-FG-02-85ER 40231.


\begin{thebibliography}{10}
\bibitem{inac} A characteristic set of recent references is given
in Ref. \cite{2FJS07} above.
\end{thebibliography}
\end{document}